\documentclass[amsmath,amssymb,aps,prl,twocolumn,
groupedaddress,showpacs]{revtex4}

\usepackage{graphicx}

\begin{document}

\title{Millisecond pulsars with r-modes\\
as steady gravitational radiators}


\author{Andreas Reisenegger and Axel Bona\v{c}i\'{c}}
\affiliation{Departamento de Astronom\'\i a y Astrof\'\i sica,
Facultad de F\'\i sica, Pontificia Universidad Cat\'olica de
Chile, Casilla 306, Santiago 22, Chile}
\email{areisene,abonacic@astro.puc.cl}

\date{\today}

\begin{abstract}

Millisecond pulsars (MSPs) are generally agreed to originate in
low-mass X-ray binaries (LMXBs), in which mass transfer onto the
neutron stars spins them up to their observed, fast rotation. The
lack of MSPs and LMXBs rotating near break-up and the similar
rotation periods of several LMXBs have been attributed to the
accretion torque being balanced, at fast rotation, by
gravitational radiation, perhaps caused by an oscillation mode
made unstable through the so-called Chandrasekhar-Friedman-Schutz
mechanism. Recently, Wagoner has argued that internal dissipation
through weak interaction processes involving $\Lambda^0$ and
$\Sigma^-$ hyperons may cause LMXBs to evolve into a quasi-steady
state, in which the neutron star has a nearly constant rotation
rate, temperature, and mode amplitude. We take this hypothesis one
step further, showing that MSPs descending from these LMXBs spend
a long time in a similar state, in which a low-amplitude r-mode
turns them into extremely steady sources of both gravitational
waves and thermal X-rays,
while they spin down due to a combination of gravitational
radiation and the standard magnetic torque. Observed MSP braking
torques already place meaningful constraints on the allowed
gravitational wave amplitudes and dissipation mechanisms.

\end{abstract}


\maketitle

The Chandrasekhar-Friedman-Schutz instability \cite{Chandra, FS1,
FS2} can make retrograde r-modes (inertial modes due to the
Coriolis force) on rapidly rotating neutron stars grow and emit
gravitational waves at the expense of the stellar rotational
energy and angular momentum \cite{Andersson98, FM}. It is choked
\cite{Lindblom98} at very high temperatures by bulk viscosity due
to non-equilibrium weak interactions such as the Urca processes
($n\to p+e+\bar\nu_e$ and $p+e\to n+\nu_e$, where $n$, $p$, $e$,
$\bar\nu_e$, and $\nu_e$ denote neutrons, protons, electrons,
electron antineutrinos, and electron neutrinos, respectively,
or analogous processes involving hyperons),
and at very low temperatures by standard shear viscosity.
Additional damping could be provided at intermediate temperatures
by hyperon bulk viscosity \cite{Jones, LindblomOwen}, caused by
the processes $\Lambda^0+n\leftrightarrow n+n$ and
$\Sigma^-+p\leftrightarrow n+n$. If strong enough, this can split
the instability region on the temperature-frequency plane into two
separate windows, as illustrated in Fig. 1.

\begin{figure}
\includegraphics[width=8.6cm,angle=0]{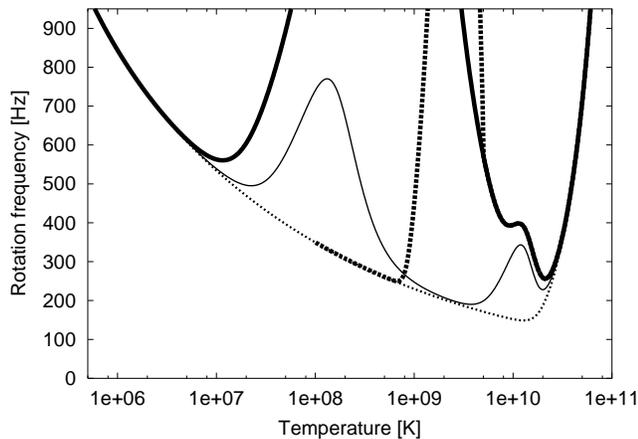}
\caption{\label{fig:fig1} Instability boundaries for the $l=m=2$
r-mode. The curves all use the same shear viscosity, determined by
Levin \& Ushomirsky \cite{Levin-b} to be active at the boundary
between the liquid core and a thin, elastic crust, but different
assumptions about the bulk viscosity in the core. The thin dotted
curve considers only modified Urca processes, whereas the other
three also include direct Urca processes and the hyperon bulk
viscosities proposed by Jones \cite{Jones} (thin solid line) and
by Lindblom \& Owen \cite{LindblomOwen} (thick solid line), in
both cases ignoring superfluid effects, and by Lindblom \& Owen
under the effects of hyperon superfluidity with a uniform, high
critical temperature, $T_c\sim5\times10^9$ K (thick dotted line).
In each case, the unstable region lies above the curve.}
\end{figure}

In the standard ``recycling'' model, millisecond pulsars (MSPs)
reach their high rotation rates due to the accretion torque
in a low-mass X-ray binary system (LMXB) \cite{Bhattacharya}. For
this to happen, some mechanism must suppress the CFS instability
enough to prevent the gravitational wave torque from being larger
than the accretion torque. The simplest scenario \cite{Levin-a}
requires
a shear viscosity substantially higher than that used in
Fig. 1, so that the instability boundary at the equilibrium
temperature $T_{acc}$ reached during the accretion process
\cite{BB} lies higher than the fastest rotation frequencies
observed in MSPs, $\sim 650$ Hz.
This is not
unlikely, since a stronger effective shear viscosity might
be obtained
with a thicker crust \cite{Levin-b},
superfluids \cite{LindblomMendell}, or a magnetic field
\cite{Mendell}. An instability boundary just above the frequency
of the fastest pulsars might even explain why there are no
MSPs approaching the break-up frequency, but it would tend to
predict an accumulation of MSPs near this boundary, rather than
the observed scarcity of MSPs with periods below 2 ms
\cite{Bildsten02}.

Wagoner \cite{Wagoner02} has shown that fast rotations
periods can
be reached even if there is a substantial unstable region at
$T_{acc}$.
In this scenario, the low-temperature instability window is
reached from below as the neutron star is spun up by accretion
(see Fig. 2(a)).
A stable balance between cooling through neutrino emission and
heating through nuclear reactions undergone by the accreted matter
in the deep crust (which release $\sim$ 1 MeV per baryon
\cite{BB}) keeps the stellar interior warm, at a temperature
$T_{acc}\sim10^{7.5-8.8}$ K that depends on the allowed neutrino
emission processes (e.g., direct vs. modified Urca). As the star
moves into the unstable region, its most unstable r-mode
\cite{Lindblom98} (with spherical-harmonic indices $l=m=2$) is
excited, producing viscous dissipation.
The resulting heat release moves the star to the high-temperature
boundary of the instability window \cite{Wagoner02, negslope}, set
by

\begin{equation}
0.051\nu_{\rm kHz}^6\ {\rm s}^{-1}=0.13f_h\nu_{\rm kHz}^2T_8^2\
{\rm s}^{-1}.
\end{equation}

\noindent The left-hand side gives the r-mode driving rate through
gravitational radiation reaction \cite{Lindblom98} for a
``fiducial'' neutron star with a polytropic ($n=1$) equation of
state, mass $M=1.4\,M_\odot$, and radius $R=12.5$ km, in terms of
the rotation frequency in kHz, $\nu_{\rm kHz}$. The right-hand
side is the damping rate, dominated by hyperonic processes, whose
large uncertainty (including the reduction due to superfluid
effects) is parameterized by the dimensionless parameter $f_h$.
This parameter is unity for the preferred bulk viscosity of
Lindblom and Owen \cite{LindblomOwen} with no superfluid effects,
acting in a hyperonic
core of radius $r=R/2$, whereas $f_h=0.16$ corresponds to the bulk
viscosity previously proposed by Jones \cite{Jones}. The resulting
interior temperature, in units of $10^8$ K, turns out to be
$T_8=0.63f_h^{-1/2}\nu_{\rm kHz}^2$.

\begin{figure}
\includegraphics[width=8.6cm,angle=0]{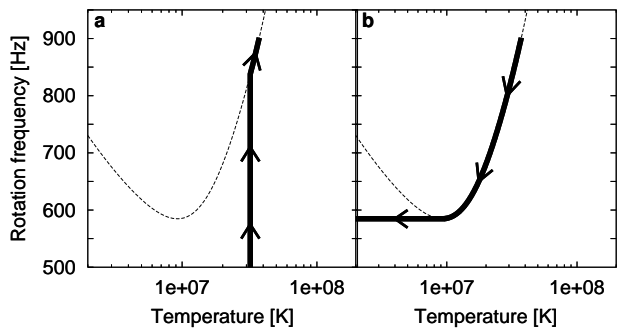}
\caption{\label{fig:fig2} Evolutionary tracks of LMXBs and MSPs.
Panel (a) shows the evolution of a neutron star in a LMXB being
spun up by accretion\cite{Wagoner02}, and panel (b) that of a MSP
being spun down by gravitational radiation and/or a magnetic
torque (solid lines). The instability boundary (dashed) considers
the same shear viscosity as in Fig. 1, and the hyperon bulk
viscosity normalized to fit the parameters of PSR B1957+20 (see
text).}
\end{figure}

Once the temperature is high enough for accretion heating to be
negligible, the dimensionless mode amplitude \cite{Lindblom98}
$\alpha$
is set by the balance of the hyperon bulk viscosity heating and
the neutrino emission (the latter assumed to be dominated by
direct Urca processes \cite{Lattimer}),

\begin{equation}
3.6\times 10^{50}f_hT_8^2\nu_{\rm kHz}^4\alpha^2{\rm erg\ s}^{-1}
=1.2\times 10^{39}f_UT_8^6{\rm erg\ s}^{-1},
\end{equation}

\noindent where we again parameterize our ignorance by a
dimensionless factor, $f_U$, which is unity when the threshold for
nucleonic direct Urca processes ($n\leftrightarrow p+e^-$) is
reached at half the stellar radius. It can take different
(possibly temperature-dependent) values if the threshold radius is
different, if only hyperonic direct Urca processes are allowed
\cite{Prakash}, or if superfluidity reduces the reaction rates
\cite{Levenfish}. The resulting amplitude is quite small,
$\alpha=7.5\times 10^{-7}(f_U/f_h^3)^{1/2}\nu_{\rm kHz}^2$, so the
linear approximation used in the expressions above are amply
justified, and it appears unlikely that the mode could be
saturated at lower amplitudes by any other processes such as
turbulent viscosity \cite{Wu} or mode coupling \cite{Arras}.

With the temperature and mode amplitude at any given rotation rate
set by eqs. (1) and (2), the accretion torque, $\dot
M\sqrt{GMR}=1.5\times 10^{34}(\dot M/\dot M_{\rm Edd})\ {\rm dyne\
cm}$ (where $\dot M$ is the mass accretion rate and $\dot M_{\rm
Edd}=1.9\times 10^{-8}M_\odot {\rm yr}^{-1}$ is the standard
``Eddington limit'' \cite{Shapiro}), gradually drives the star
``up'' towards faster rotation. It may eventually settle in a
quasi-steady state, in which the accretion torque is balanced by
the angular momentum loss due to gravitational radiation
\cite{Wagoner02}, $6.8\times 10^{46}\nu_{\rm kHz}^7\alpha^2\ {\rm
dyne\ cm}$, at an equilibrium period

\begin{equation}
P_{\rm eq}={1\over\nu_{\rm eq}}=1.1\left({f_U\over f_h^3}{\dot
M_{\rm Edd}\over\dot M}\right)^{1/11}{\rm ms},
\end{equation}

\noindent which is only very weakly dependent on the uncertain
parameters. In this state, the neutron star is a source of
low-amplitude gravitational waves, whose steadiness depends on
that of the accretion rate.

It is worth noting that eq. (3) and the expression above for the
accretion torque are only valid if the star's rotation is much
slower than the mass-shedding limit and the accretion disk extends
down to the stellar surface, not being constrained by the star's
magnetic field. We also point out that the possibility of
accretion torques in LMXBs being balanced by gravitational wave
emission, in somewhat different scenarios, had previously been
suggested \cite{Wagoner84, Bildsten98}.

In what follows, we analyze
what happens in Wagoner's scenario once the accretion stops and
the star turns on as a (millisecond) radio pulsar. It slowly
slides down the instability boundary as it loses angular momentum
through the combined effects of gravitational radiation and
standard magnetic braking (Fig. 2(b)). Its heating and cooling
time scales, as well as the growth and damping times for the
r-modes, are much shorter than the spin-down time, and therefore
the star will remain in the quasi-steady state described by
equations (1) and (2). Since no external perturbations are
present, the spin-down is extremely smooth and predictable, as
observed in MSPs \cite{Lorimer}. For a given neutron star model
(which sets the ``fudge factors'' $f_U$ and $f_h$), the rotation
period, $P$ (observed as the periodicity of the radio pulses),
uniquely determines the gravitational-wave contribution to the
spin-down rate

\begin{equation}
\dot P_{GW}=5.3\times 10^{-18}{f_U\over f_h^3}\left(1{\rm ms}\over
P\right)^9.
\end{equation}

\noindent We note that, if the particles in the neutron star core
are superfluid, the factors $f_U$ and $f_h$ will depend on
temperature, and therefore implicitly on $P$, which would modify
the dependence of $\dot P_{GW}$ from the proportionality to
$P^{-9}$ implied by eq. (4).
However, for typical cases that we evaluated numerically,
a strongly decreasing function $\dot P_{GW}(P)$ is still obtained.

Of course, for periods $P$ of actually observed MSPs in the range
of applicability of this model, $\dot P_{GW}(P)$ can be at most as
high as the observed $\dot P$.
The most constraining case
is the so-called ``black widow'' \cite{Fruchter}, PSR B1957+20,
because of its short period, $P=1.60$ ms, and its small intrinsic
period derivative \cite{Camilo}, $\dot P=1.2\times 10^{-20}$. For
this pulsar, we require $f_h^3/f_U\geq 6.4,$ i.e., a somewhat
larger bulk viscosity and/or a substantially lower neutrino
emissivity than in our favored model. Neither of these would be
too surprising, in view of their strong dependence on the
uncertain state of very dense matter. The presence of superfluid
energy gaps for the particles in the neutron star core could
substantially reduce the phase space and therefore the rates of
all reactions, thereby decreasing both the Urca cooling rate
\cite{Levenfish} and the hyperon bulk viscosity
\cite{LindblomOwen, Haensel02}, potentially down to $f_U,f_h\ll
1$. Since the reactions determining $f_h$ involve more potentially
superfluid particles (both hyperons and nucleons) than the Urca
processes (just nucleons),
it is likely that
$f_h < f_U$, making it difficult to satisfy the ``black widow''
constraint. However, $f_U$ may be reduced if, instead of nucleonic
direct Urca processes, only hyperonic direct Urca reactions are
allowed, which would set $f_U$ (without superfluid reduction) at
$\sim 0.2$ \cite{Prakash}. The latter processes are almost
certainly allowed as long as hyperons exist in the star, but again
may be reduced (perhaps more strongly than the neutrino-less
processes contributing to $f_h$) if the hyperons become superfluid
at high temperatures.
Therefore, instead of
ruling out the model, the observations may be used to constrain
the internal properties of neutron stars once the model is
validated in other ways.

Of course, the values of $f_h$ and $f_U$
are most likely not identical for all MSPs, as these probably have
different masses, and therefore different mean densities and
different fractions of their interior in which the relevant
processes take place. Assuming, for simplicity, that the condition
$f_h^3/f_U\geq 6.4$
does apply to all MSPs and LMXBs, we can in principle constrain
the LMXB equilibrium period to $P_{\rm eq}\leq 0.9(\dot M_{\rm
Edd}/\dot M)^{1/11}$ ms. Strictly speaking, this period is beyond
the mass-shedding limit for our adopted neutron star model, and
should therefore not be taken literally. It shows that, in order
to be consistent with MSP observations, the r-mode instability
cannot limit the rotation of LMXBs to much below the maximum
allowed equilibrium rotation for any neutron star, unless
magnetic stresses prevent the accretion disk from reaching the
surface,
reducing the accretion torque and increasing the
equilibrium period. A similar conclusion has been reached before
\cite{BU}, based on the low luminosity of LMXBs in quiescence,
compared to the expectation from internal dissipation.

The spin-down of a newly born MSP, initially spinning at a period
$P_0$ near the mass-shedding limit, is likely to be dominated by
the gravitational radiation torque. The time it takes to spin down
to a slower period $P$ is $t\approx 3.9\times
10^6(f_h^3/6.4f_U)[(P/{\rm ms})^{10}-(P_0/{\rm ms})^{10}]$ yr,
substantially shorter (and much more period-dependent) than for
magnetic braking at the inferred magnetic field strengths, $\sim
10^{8-9}$ G. This makes it much less likely to find such extremely
rapidly spinning MSPs, perhaps in this way explaining the observed
absence of these objects \cite{Bildsten02}. An independent test of
this scenario would be a measurement of the braking index,
$n\equiv \nu\ddot\nu/\dot\nu^2$ \cite{Shapiro}, which reaches a
huge value ($n=11$, if $f_h^3/f_U$ is constant) in the regime
where the torque is dominated by gravitational-wave emission
(compared to $n=3$ for pure magnetic dipole braking).
Unfortunately, measurements of braking indices in MSPs are
out of reach.

In this state, the effective surface temperature of the MSP can be
inferred from its interior temperature \cite{Gudmundsson} to be
$T_s\approx 8\times 10^5f_h^{-0.28}(P/{\rm ms})^{-1.1}$ K, not far
below the upper limits obtained from ROSAT observations of MSPs
\cite{Reisenegger97}, and perhaps within the reach of more careful
determinations based on XMM-Newton spectra.

Perhaps most interestingly, the MSP will radiate gravitational
waves at the r-mode frequency. In the limit of slow rotation and
weak gravity, this frequency is 4/3 times the rotation frequency
\cite{Papaloizou}, or $\approx 850$ Hz for the fastest pulsars. A
more precise determination, including centrifugal and relativistic
corrections, should be possible.
The angle-averaged amplitude at a distance $D$
is \cite{Owen}

\begin{equation}
h\approx\ 7\times 10^{-28}\left(6.4f_U\over
f_h^3\right)^{1/2}\left(1.6{\rm ms}\over P\right)^{5}{1.5{\rm
kpc}\over D},
\end{equation}

\noindent where the reference numbers correspond to the ``black
widow'' pulsar, the most favorable known so far. According to
recent sensitivity curves \cite{Brady}, this signal is almost
within reach of the advanced LIGO with signal recycling, tuned at
the appropriate frequency and integrating for $1/3$ yr. Therefore,
future gravitational-wave observatories may well give us
information about weak interaction processes in superdense matter.

A recent study \cite{Andersson02} has found that ``strange
stars'', composed of ``deconfined'' $u$, $d$, and $s$ quarks, may
have a time evolution qualitatively similar to that later found by
Wagoner \cite{Wagoner02} for neutron stars in LMXBs and here for
neutron stars as MSPs. In the corresponding strange stars, the
bulk viscosity is dominated by the process $u+d\leftrightarrow
s+u$, the quark analog for the hyperonic processes considered in
the present discussion. In that study, the reheating due to bulk
viscous dissipation is ignored, and therefore the evolution time
scale is set by passive cooling, which is much faster and thus
more violent than in the more realistic case discussed here. In
view of the present theoretical uncertainties and lack of
observational constraints, LMXBs and MSPs may well be strange
stars rather than neutron stars. Although the numerical details
are different (and even more uncertain), ``strange MSPs'' may
follow a similar evolutionary path and be subject to observational
constraints analogous to those discussed here for neutron stars.

We conclude that, if there is a substantial instability window at
low temperatures, neutron stars (or strange stars) in LMXBs can
generally spin up to MSP rotation rates by the mechanism suggested
by Wagoner \cite{Wagoner02}.
The gravitational wave torque
will still be active
in their MSP phase, until they reach the bottom of the instability
window, which some of the observed MSPs may not yet have done.
Thus, these MSPs will be
sources of gravitational waves
and thermal X-rays.
Even if born rotating near break-up,
MSPs will spin down to near the observed periods much more quickly
than in the magnetic braking model, perhaps explaining the
scarcity of very fast MSPs.

\begin{acknowledgments}

The authors thank T. Creighton and L. Lindblom for useful
information, N. Andersson, P. Arras, M. Catelan, Y. Wu, and two
anonymous referees for comments that improved the manuscript, and
L. Bildsten for encouragement. This work was supported by
FONDECYT-Chile through the regular research grant \# 1020840.
\end{acknowledgments}


\begin{thebibliography}{}
\bibitem[Bonacic (2003)]{negslope} Our numerical calculations show that, even if the
        instability boundary is reached on its
        low-temperature side (where $d\nu/dT<0$), the heating due to
        shear viscosity and hyperon bulk viscosity combined is strong
        enough to move the star across the window to the high-temperature
        side, so the present analysis still applies.
\end{thebibliography}
\end{document}